\documentclass[twocolumn,a4paper]{revtex4-1}
\usepackage{hyperref}
\usepackage{graphicx}
\usepackage{amsmath}
\usepackage{xcolor}
\usepackage{float}
\usepackage{amsfonts}
\usepackage{epsfig}
\usepackage{graphics}
\usepackage{color}
\usepackage{csquotes}
\newcommand{\be}{\begin{equation}}
\newcommand{\ee}{\end{equation}}
\newcommand{\ba}{\begin{eqnarray}}
\newcommand{\ea}{\end{eqnarray}}

\newcommand{\beq}{\begin{equation}}
\newcommand{\eeq}{\end{equation}}
\newcommand{\bqa}{\begin{eqnarray}}
\newcommand{\eqa}{\end{eqnarray}}

\begin{document}
\title{Dissociation of J/$\psi$ and $\Upsilon$ using dissociation energy criteria in N-dimensional space}
\author{Siddhartha Solanki$^{a}$}
\author{Manohar Lal$^{a}$} 
\author{Vineet Kumar Agotiya$^{a}$}
\email{agotiya81@gmail.com}
\affiliation{$^a$Department of Physics, Central University of Jharkhand, Ranchi, India, 835222}

\begin{abstract}
The analytical exact iteration method (AEIM) have been used widely to calculate N-dimensional radial Schrodinger equation with medium modified form of Cornell potential and is generalized to the finite value of magnetic field (eB) with quasi-particle approach in hot quantum chromodynamics (QCD) medium. In N-dimensional space the energy eigen values have been calculated for any states (n,l). These results have been used to study the properties of quarkonium states (i.e, the binding energy and mass spectra, dissociation temperature and thermodynamical properties in the N-dimensional space). We have determined the binding energy of the ground states of quarkonium with magnetic field and dimensionality number. We have also determined the effects of magnetic field and dimensionality number on mass spectra for ground states of quarkonia. But main result is quite noticeable for the values of dissociation temperature in terms of magnetic field and dimensionality number for ground states of quarkonia after using the criteria of dissociation energy. At last, we have also calculated the thermodynamical properties of QGP (i.e., pressure, energy density and speed of sound) using the parameter eB with ideal equation of states (EoS).\\

\end{abstract}

\maketitle

\section{Introduction}
The system of heavy quarkonia (such as Bottomonium and Charmonium) have played a key role for the comprehensive and quantitative test of Quantum Chromo-Dynamics (QCD) and the standard model\cite{1}. After discovery of J/$\psi$ in 1974, the study of heavy quarkonia becomes interesting topic for both the theoretical and experimental high energy physicists. The radial Schrodinger equation has been solved with the real part of medium modified form of Cornell potential\cite{2}, and the solution, thus, obtained can be used to understand many phenomena in the study of atomic and molecular physics, spectroscopy (hadronic as well as molecular), nuclear physics and also in high energy physics which are not yet understood. Most of the recent quarkonium studies are focused on N-dimensional space problem\cite{3}, lower spatial or dimensional space problem\cite{4}. The consequences of N-dimensional space have been considered for energy levels of the bound state system of quarkonia\cite{5}. In the N-dimensional space, the study of the harmonic oscillator\cite{6} and hydrogen atom\cite{7} have also been done. Additionally, in N-dimensional space, the Schrodinger equation has also been solved for potentials such as Cornell potential\cite{8}, fourth order inverse power potential\cite{9}, Mie type potential\cite{10}, Kratzer potential\cite{11}, Coulomb potential\cite{12}, Energy dependent potential\cite{13}, Global potential\cite{14}, Hua potential\cite{13} and Harmonic potential\cite{3} etc.\\
There were several methods used to solve the Schrodinger equation such as, power series method\cite{1}, Hill determinant method\cite{15}, numerical methods\cite{16,17,18}, quasi-linearization method (QLM)\cite{19}, point canonical transformation (PCT)\cite{20} and operator algebraic method\cite{21}, Nikiforov-Uvarov (UV) method\cite{5,13,22}, power series technique\cite{23}, Laplace transformation methods\cite{3,12}, Asymptotic iteration method (AIM)\cite{2}, SUSYQM method, AEIM\cite{24} and Pekeris type approximation\cite{13,25} etc.\\
\begin{figure*}
\centering
    \vspace{3mm}   
    \includegraphics[height=7cm,width=9cm]{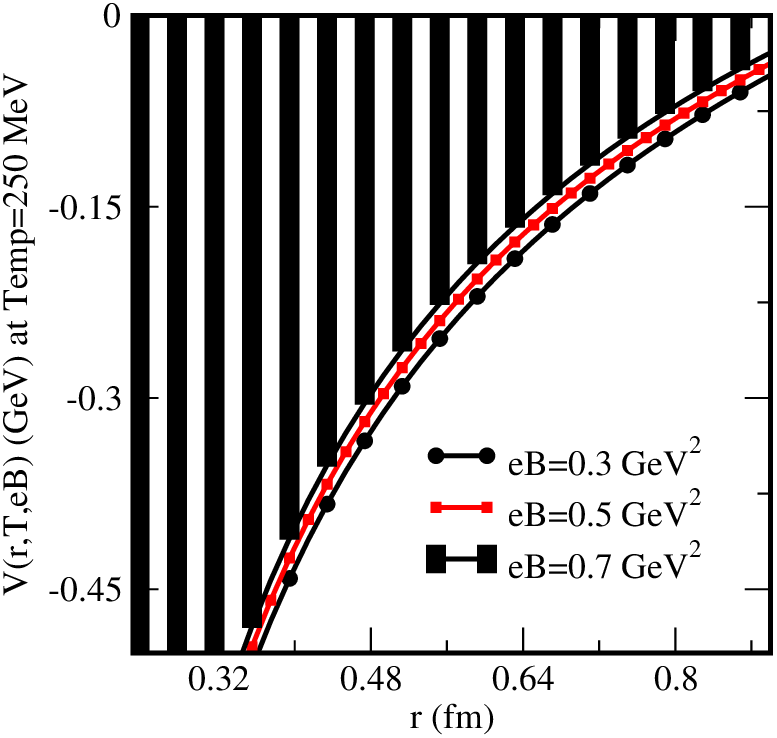}
     \vspace{1cm}  
\caption{The variation of V(r,T,eB) with distance (r) at different values of magnetic field and fixed value of T=250 MeV.}
\label{fig.1}
 \vspace{2cm}  
\end{figure*}
The quarkonium dissociation rates for the ground state have been studied by direct continuum of thermal activation, tunneling, binding energy and phase shift scattering in the hot quark-gluon plasma (QGP) for the eigen states of lowest values\cite{17,26}. Vija and Thoma\cite{27} extended perturbated gauge theory for the collisional energy loss in QGP at finite chemical potential and temperature. The quarkonium dissociation has been studied by correcting Cornell potential via hard thermal loop resumed propagator of the gluon\cite{28,29}. The study of the heavy quarkonium binding energy in details is found in\cite{30,31} and the chemical potential effect have been also studied by the methods of dissipative hydrodynamic on quark-gluon plasma, production of photon in QGP and the quark-gluon plasma thermodynamical propertie\cite{32,33,34,35,36}. The Alberico et al.\cite{37}, Mocsy et al.\cite{38} and Agotiya et al.\cite{39} have solved the Schrodinger equation for the quarkonium states at finite temperature, using a temperature dependent effective potential by the linear combination of internal energy and concluded the spectral function of quarkonium in a quark-gluon plasma.\\ 
In this work, The effect of magnetic field has been introduced through quasi-particle debye mass. We have used medium modified form of potential, so formed, obtained the binding energies and mass spectra in N-dimensional space of charmonium and bottomonium at different values of magnetic field and dimensionality number. The dissociation energy of QGP have been introduced for the calculation of dissociation temperature of the ground states of quarkonia by the intersection point of dissociation energy and binding energy in N-dimensional space. In other studies, authors have also calculated the dissociation temperature by using the criteria that at dissociation point thermal width is equal to twice of binding energy. The effect of magnetic field and dimensionality number significantly revise values of dissociation points. At last we have also calculated the thermodynamical properties of QGP (i.e., pressure, energy density and speed of sound) using the eB and dimensionality number (N) and compared with the previous published data.\\
The paper is organized as follows: In section-II, the exact solution of N-dimensional Radial Schrodinger equation with the medium modified form of Cornell potential has been calculated. In section-III, Quasi particle model and Debye mass has been discussed. In section-IV, the binding energy of quarkonium state in N-dimensional space has been investigated. In section-V, a brief description about the mass spectra of quarkonium state in N-dimensional space has been provided. In section-VI, study about the QCD EoS in presence of eB. In section-VII, the Dissociation energy (D.E.) of Quarkonium state in N-dimensional space has been investigated. Results has been discussed in section-VIII and work has been concluded in section-IX. 
\begin{figure*}
    \vspace{1cm}   
    \includegraphics[height=6cm,width=8cm]{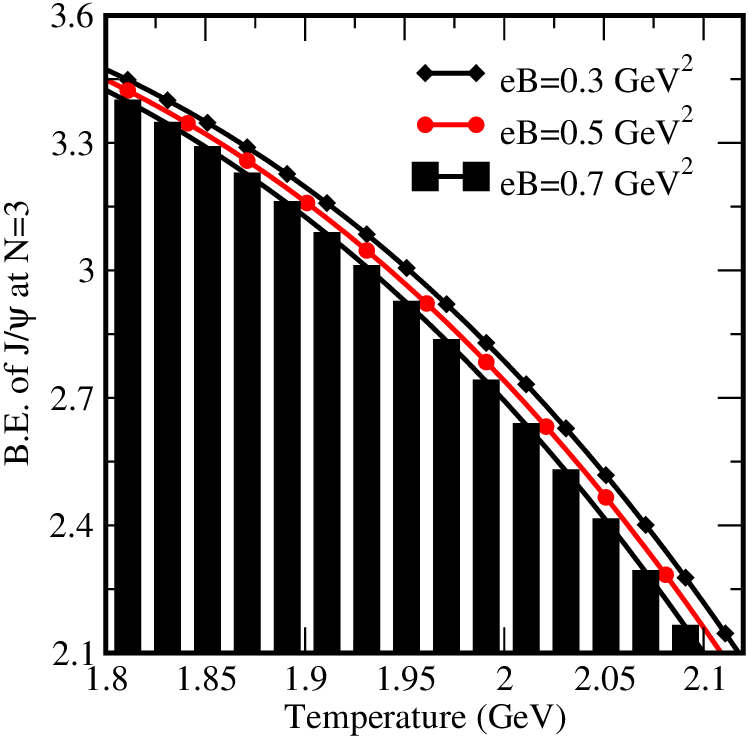}
    \hspace{8mm}
    \includegraphics[height=6cm,width=8cm]{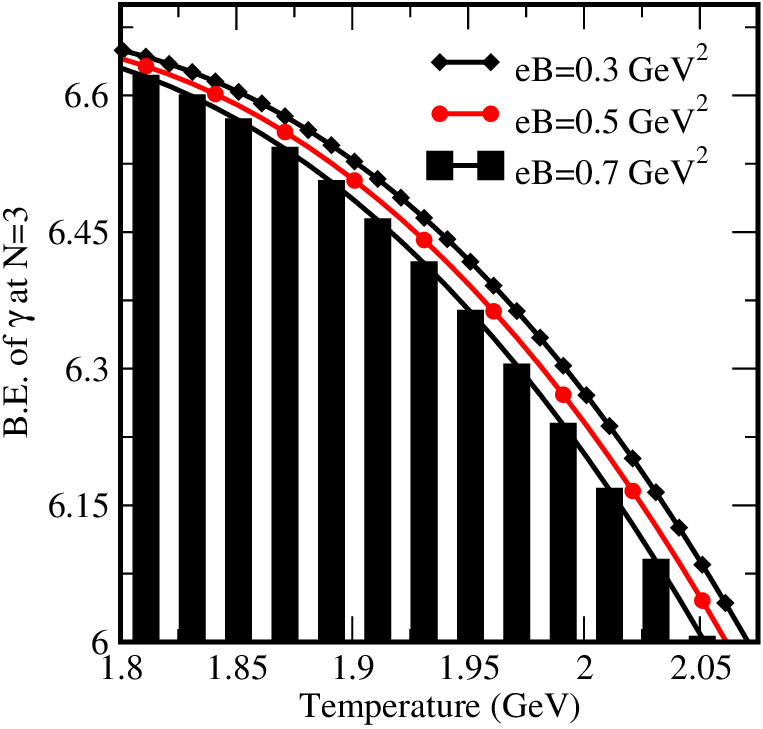}
    \vspace{2cm} 
\caption{Dependence of J/$\psi$ (left panel) and $\Upsilon$ (right panel) binding energy with temperature for different values of magnetic field at fixed value of N=3.}
\label{fig.2}
 \vspace{2cm} 
\end{figure*}

\section{The solution of N-dimensional Radial Schrodinger equation with the medium modified form of Cornell potential}
The N-dimensional radial Schrodinger equation for two interacting particles has been solved for the medium modified form of the Cornell potential using AEIM method\cite{40,41}
\begin{equation}
\label{eq1}
\frac{\mathrm{d}^2}{\mathrm{d} r^2}+\frac{N-1}{r}\frac{\mathrm{d} }{\mathrm{d} r}-\frac{l(l+N-2)}{r^2}+2\mu_{Q\bar{Q}}\left[E_{nl}-V(r)\right]\psi(r)=0
\end{equation}
Where, N, l and $\mu_{Q\bar{Q}}$ are the dimensional number, angular quantum number and reduced mass of bound state respectively. Now, the wave function [$\psi (r)$] choosen here is,
\begin{equation}
\label{eq2} 
\psi (r)= R(r)/r^{\frac{N-1}{2}} 
\end{equation}
Putting Eq.(\ref{eq2}) in Eq.(\ref{eq1}), we get
\begin{equation}
\label{eq3}
\left [\frac{\mathrm{d}^2}{\mathrm{d} r^2}-\frac{\lambda ^2-\frac{1}{4}}{r^2}+2\mu_{Q\bar{Q}}\left [E_{nl}-V(r) \right ] \right ]R(r)=0
\end{equation}
\begin{figure*}
    \vspace{1cm}   
    \includegraphics[height=6cm,width=8cm]{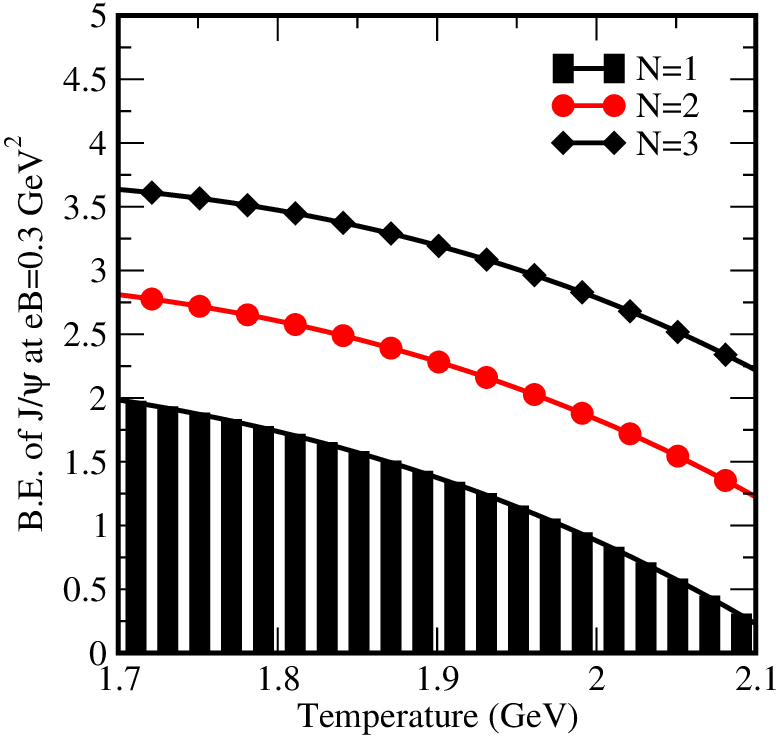}
    \hspace{8mm}
    \includegraphics[height=6cm,width=8cm]{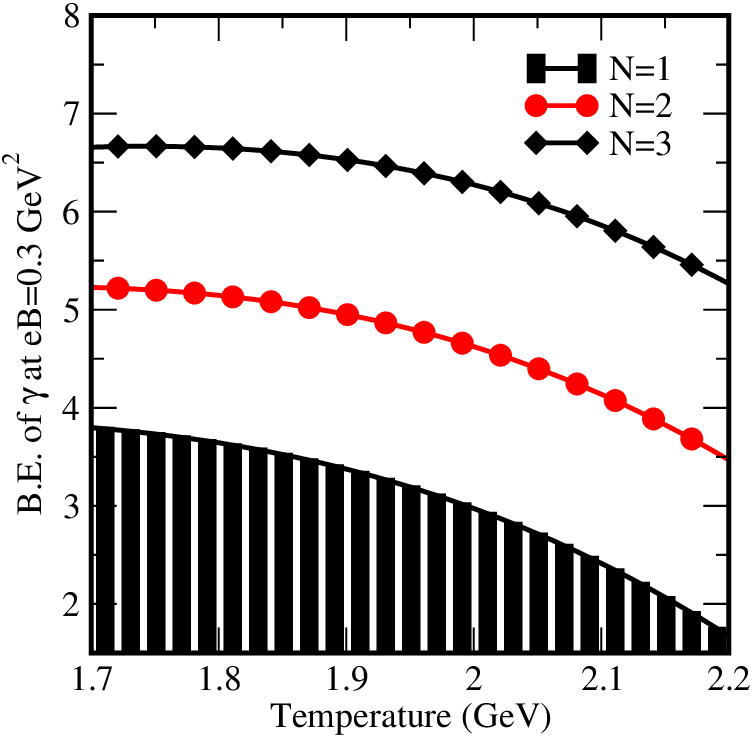}
    \vspace{1cm} 
\caption{Dependence of J/$\psi$ (left panel) and $\Upsilon$ (right panel) binding energy with temperature for different values of dimensionality number at fixed value of eB=0.3 Ge$V^{2}$.}
\label{fig.3}
 \vspace{3cm} 
\end{figure*}
Where,
\begin{equation}
\label{eq4} 
\lambda =l+\left(\frac{N-2}{2}\right)
\end{equation}
We have taken the spatially isotropic form of heavy quark cornell potential to obtain the in-medium modification i.e.,
\begin{equation}
\label{eq5}
V(r)=-\frac{\alpha}{r}+\sigma r
\end{equation}
In general, the spacial form of the potential may have anisotropic structure due to the breaking of spherical symmetry in the presence of magnetic field\cite{Iwasaki,Nilima23}. In our work, the in-medium modification of the above equation Eq.(\ref{eq5}) have been obtained in the momentum space by dividing the vacuum heavy quark potential with the medium dielectric permittivity\cite{Nilima23}, which carries the information of temperature and magnetic field (eB). In that case, we have found the in-medium modified form of cornell potential\cite{42} for the study of quarkonia in the presence of eB is:  
\begin{equation}
\label{eq6}
V(r)=\frac{\sigma}{m_{D}}\left[1-exp(-m_{D}r)\right]-\frac{\alpha}{r}[exp(-m_{D}r)]
\end{equation}
Using exponential formula $e^{-m_{D}r} = \sum_{k=0 }^{\infty }\frac{(-m_{D}r)^{k}}{k!}$ for solving the Eq.(\ref{eq5}) and neglecting the
higher orders at $m_{D}r\ll1$, thus Eq.(\ref{eq6}) takes the following form,
\begin{equation}
\label{eq7}
V(r)=-ar^{2}+br+c-\frac{d}{r}
\end{equation}
Where, the values of $a$, $b$, $c$ and $d$ are given as $a=\frac{1}{2}(\sigma m_{D})$, $b=\frac{1}{2}(2\sigma-\alpha m_{D}^{2})$, $c=\alpha m_{D}$ and $d=\alpha$, substituting these values in Eq.(\ref{eq3}), we get the radial wave function:
\begin{equation}
\label{eq8}
{R}''(r)=\left [ -\varepsilon _{nl}-a_{1}r^{2}+b_{1}r+c_{1}-\frac{d_{1}}{r}+\frac{\lambda^{2}-\frac{1}{4}}{r^{2}} \right ]R(r)
\end{equation}
Where, $\varepsilon _{nl}=2\mu_{Q\bar{Q}}E_{nl}$, $a_{1}=|-2\mu_{Q\bar{Q}}a|$, $b_{1}=2\mu_{Q\bar{Q}}b$, $c=2\mu_{Q\bar{Q}}c$ and $d_{1}=2\mu_{Q\bar{Q}}d$. AEIM requires making the following ansatz for the wave function as in\cite{11,43,44}.
\begin{equation}
\label{eq9}
R(r)=f_{n}(r)exp[g_{1}(r)]
\end{equation}
Where,
\begin{equation}
\label{eq10}
f_{n}(r)=1 
\end{equation}if n=0 
and
\begin{equation}
\label{eq11}
f_{n}(r)=\prod_{i=1}^{n}\left ( r-\alpha_{i}^{(n)}  \right ) 
\end{equation} for n=1, 2, 3........
\begin{equation}
\label{eq12}
g_{1}(r)=-\frac{1}{2}\alpha r^{2}-\beta r+\delta lnr , \alpha >0, \beta >0
\end{equation}
From Eq.(\ref{eq8}), we get
\begin{equation}
\label{eq13}
{R}''_{n,l}(r)=\left ( {g}''_{l}(r)+ g_{l}^{'2}(r)+\frac{{f}''_{n}(r)+2g_{l}^{'}(r)f_{n}^{'}(r)}{f_{n}(r)} \right )R_{nl}(r)
\end{equation}
\begin{table}
\label{tab1}
\centering
\caption{Dissociation of lower bound for eB=0.3 $GeV^{2}$, temperatures are in the unit of $T_c$ by using thermal energy effect criteria.}
{\begin{tabular}{@{}cccc@{}} \toprule
$State$$\Downarrow$ & N=3 & N=4 & N=5\\ \colrule
$J/\psi$ & 1.3921 & 1.5109 & 1.7709\\ \colrule
$\Upsilon$ & 2.0509 & 2.2509 & 2.4110\\ \botrule
\end{tabular}}
\end{table}
\begin{figure*}
    \vspace{1mm}   
    \includegraphics[height=6cm,width=8cm]{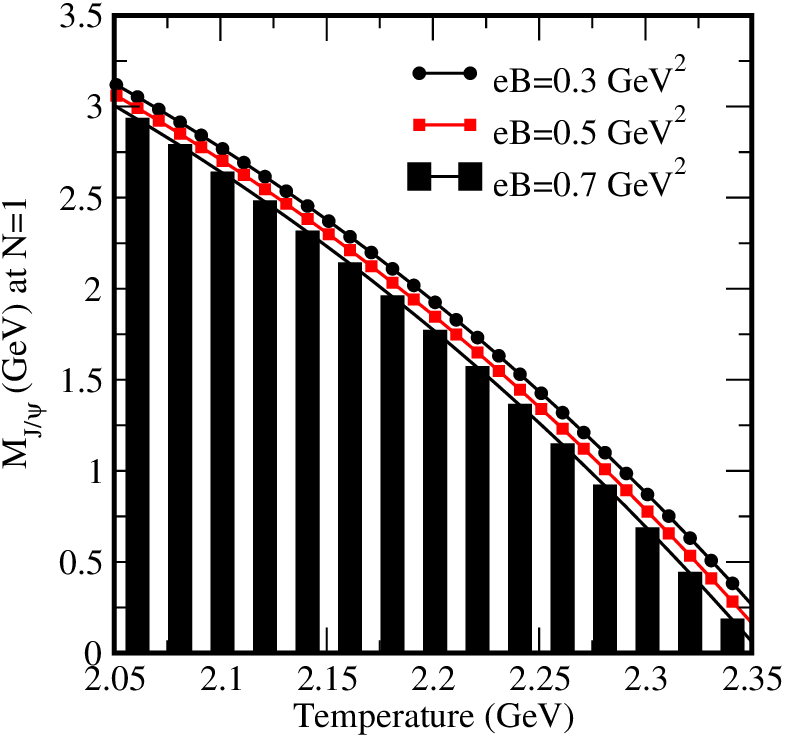}
    \hspace{8mm}
    \includegraphics[height=6cm,width=8cm]{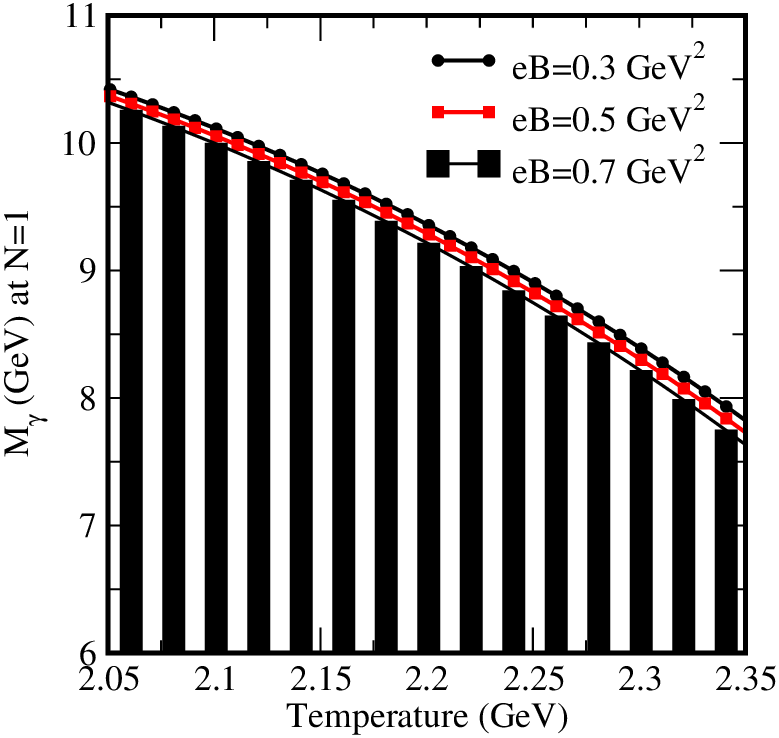}
\caption{Dependence of J/$\psi$ (left panel) and $\Upsilon$ (right panel) mass spectra with temperature for different values of magnetic field at fixed value of N=1.}
\label{fig.4}
 \vspace{2cm} 
\end{figure*}
\begin{table}
\label{tab2}
\centering
\caption{Dissociation of lower bound for N=5, temperatures are in the unit of $T_c$ by using thermal energy effect criteria.}
{\begin{tabular}{@{}cccc@{}} \toprule
$State$$\Downarrow$ & eB=0.3 $GeV^{2}$ & eB=0.5 $GeV^{2}$ & eB=0.7 $GeV^{2}$\\ \colrule
$J/\psi$ & 1.7709 & 1.7591 & 1.7465\\ \colrule
$\Upsilon$ & 2.4110 & 2.4009 & 2.3899\\ \botrule
\end{tabular}}
\end{table}
After comparing the Eq.(\ref{eq13}) and Eq.(\ref{eq9}) we have:
\begin{multline}
\label{eq14}
-\varepsilon _{n,l}+a_{1}r^{2}+b_{1}r+C_{1}-\frac{d_{1}}{r}+\frac{\lambda^{2}-\frac{1}{4} }{r^{2}}= {g}''_{l}(r)+ g_{l}^{'2}(r)\\+\frac{{f}''_{n}(r)+2g_{l}^{'}(r)f_{n}^{'}(r)}{f_{n}(r)}
\end{multline}
At n=0, substitute Eqs.(\ref{eq10}), (\ref{eq11}), (\ref{eq12}) and Eq.(\ref{eq13}) into Eq.(\ref{eq14}) we get
\begin{multline}
\label{eq15}
a_{1}r^{2}+b_{1}r+C_{1}-\frac{d_{1}}{r}+\frac{\lambda^{2}-\frac{1}{4} }{r^{2}}-\varepsilon_{0l}
= \alpha^{2}r^{2}\\+2\alpha \beta r-\alpha [1+2(\delta +0)]+\beta ^{2}-\frac{2\beta \delta }{r}+\frac{\delta (\delta -1)}{r^{2}}
\end{multline}
Now comparing the coefficient of $r$ on both side of Eq.(\ref{eq15}), we obtain
\begin{equation}
\label{eq16}
\alpha =\sqrt{a_{1}}
\end{equation}
\begin{equation}
\label{eq17}
\beta = \frac{b_{1}}{2\sqrt{a_{1}}}
\end{equation}
\begin{equation}
\label{eq18}
d_{1}=2\beta (\delta +0)
\end{equation}
\begin{equation}
\label{eq19}
\delta = \frac{1}{2}(1\pm 2\lambda )
\end{equation}
\begin{equation}
\label{eq20}
\varepsilon_{0l} =\alpha [1+2(\delta +0)]+C_{1}-\beta ^{2}
\end{equation}
Now the energy eigen value for the ground state is:
\begin{equation}
\label{eq21}
E_{0l}= \sqrt{\frac{a}{2\mu_{Q\bar{Q}}}}(N+2l)+C-\frac{b^{2}}{4(\sqrt{a})^{2}}
\end{equation}
Now, for the first node (n=$1$), we used the function $f_{1}(r)=\left ( r-\alpha_{1}^{(1)}  \right )$ and $g_{1}(r)$ then,
\begin{figure*}
    \vspace{2cm}   
    \includegraphics[height=6cm,width=8cm]{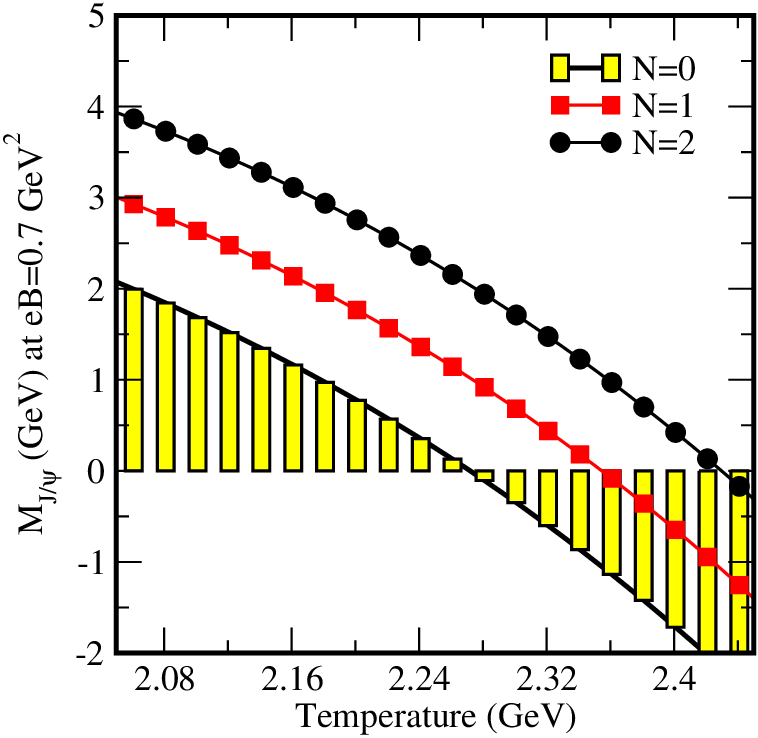}
    \hspace{8mm}
    \includegraphics[height=6cm,width=8cm]{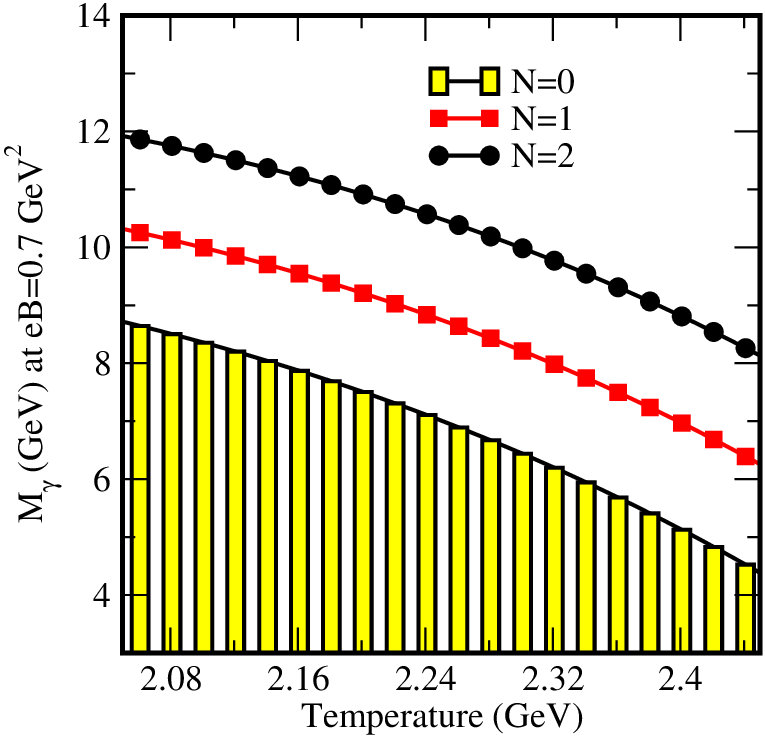}
\caption{Dependence of J/$\psi$ (left panel) and $\Upsilon$ (right panel) mass spectra with temperature for different values of dimensionality number at fixed value of eB=0.7 Ge$V^{2}$.}
\label{fig.5}
 \vspace{2cm} 
\end{figure*}
\begin{multline}
\label{eq22}
a_{1}r^{2}+b_{1}r+C_{1}-\frac{d_{1}}{r}+\frac{\lambda^{2}-\frac{1}{4} }{r^{2}}-\varepsilon_{1l} = 
\alpha^{2}r^{2}\\+2\alpha \beta r-\alpha [1+2(\delta +1)]+\beta^{2}-\frac{2[\beta (\delta +1)
+\alpha \alpha_{1}^{(1)} ]}{r}+\frac{\delta (\delta -1)}{r^{2}}
\end{multline}
By comparing the coefficients of $r$, the relation between the potential parameters are:
\begin{equation}
\label{eq23}
\alpha =\sqrt{a_{1}}
\end{equation}
\begin{equation}
\label{eq24}
\beta =\frac{b_{1}}{2\sqrt{a_{1}}}
\end{equation}
\begin{equation}
\label{eq25}
d_{1}=2\beta (\delta +1)
\end{equation}
\begin{equation}
\label{eq26}
\delta =\frac{1}{2}(1\pm 2\lambda )
\end{equation}
\begin{equation}
\label{eq27}
\varepsilon_{1l} = \alpha [1+2(\delta +1)]+C_{1}-\beta^{2}
\end{equation}
\begin{equation}
\label{eq28} 
d_{1}-2\beta (\delta +1)=2\alpha \alpha_{1}^{(1)}
\end{equation}
\begin{equation}
\label{eq29}
(d_{1}-2\beta \delta )\alpha_{1}^{(1)} =2\delta 
\end{equation}
Now, the energy eigen value formula for first excited state $E_{1l}$ is:
\begin{equation}
\label{eq30}
E_{1l}=\sqrt{\frac{a}{2\mu_{Q\bar{Q}}}}(N+2l+2)+C-\frac{b^{2}}{4(\sqrt{a})^{2}}
\end{equation}
Similarly, for second node $(n=2)$, we use $f_{2}(r)=\left ( r-\alpha _{1}^{(2)} \right ) \left ( r-\alpha _{2}^{(2)} \right )$ and $g_{1}(r)$
and we get:
\begin{multline}
\label{eq31}
a_{1}r^{2}+b_{1}r+C_{1}-\frac{d_{1}}{r}+\frac{\lambda^{2}-\frac{1}{4} }{r^{2}}-\varepsilon _{2l}=\alpha^{2} r^{2}+2\alpha \beta r-\\\alpha [1+2(\delta +2)]\\+\beta ^{2}-\frac{2[\beta (\delta +2)+\alpha (\alpha_{1}^{(2)}+\alpha_{2}^{(2)})]}{r}+\frac{\delta (\delta -1)}{r^{2}}
\end{multline}
By comparing the coefficients of r we get:
\begin{equation}
\label{eq32}
\alpha =\sqrt{a_{1}}
\end{equation}
\begin{equation}
\label{eq33}
\beta =\frac{b_{1}}{2\sqrt{a_{1}}}
\end{equation}
\begin{equation}
\label{eq34}
\delta =\frac{1}{2}(1\pm 2\lambda )
\end{equation}
\begin{equation}
\label{eq35}
\varepsilon_{2l}=\alpha [1+2(\delta +2)]+C_{1}-\beta^{2}
\end{equation}
\begin{figure*}
\centering
    \vspace{3mm}   
    \includegraphics[height=8cm,width=14cm]{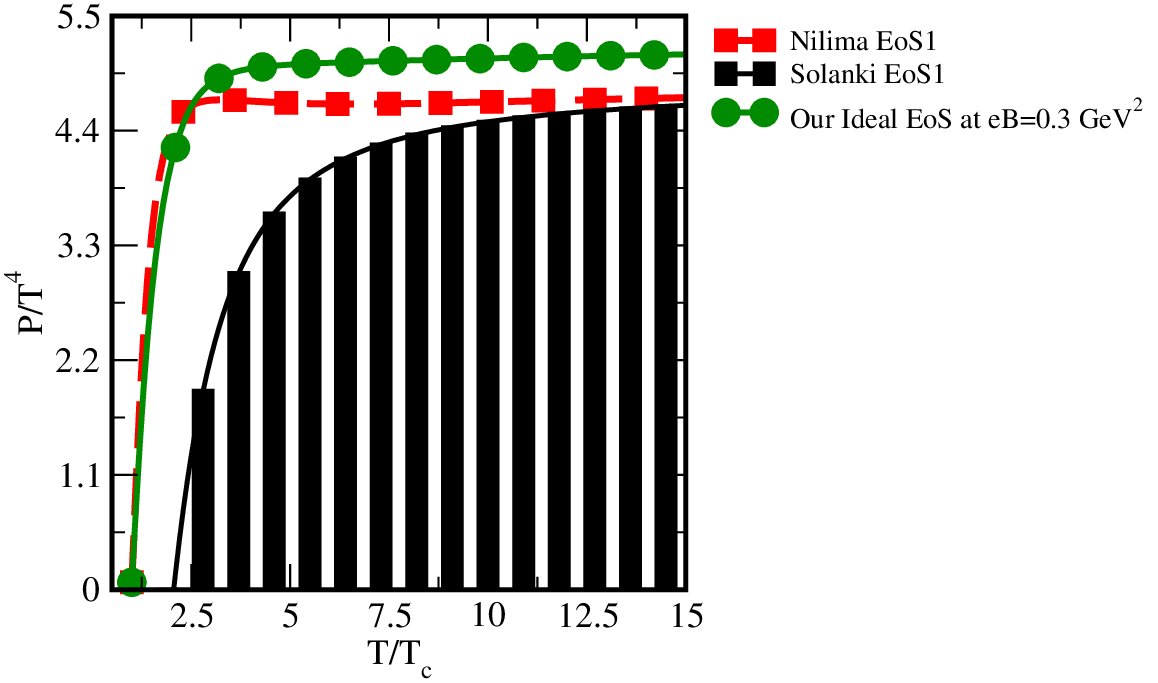}
\vspace{2cm}
\caption{Variation of $\frac{P}{T^{4}}$ as a function of $\frac{T}{T_{c}}$ for our ideal EoS at eB=0.3 Ge$V^{2}$ and also compared with the published results of Nilima EoS1~\cite{56} and Solanki EoS1~\cite{57}.}
\label{fig.6} 
\vspace{3cm}
\end{figure*}
\begin{table}
\label{tab3}
\centering
\caption{Dissociation of upper bound for eB=0.3 $GeV^{2}$, temperatures are in the unit of $T_c$ by using thermal energy effect criteria.}
{\begin{tabular}{@{}cccc@{}} \toprule
$State$$\Downarrow$ & N=3 & N=4 & N=5\\ \colrule
$J/\psi$ & 2.2109 & 2.3309 & 2.4310\\ \colrule
$\Upsilon$ & 2.4710 & 2.5909 & 2.7010\\ \botrule
\end{tabular}}
\end{table}
\begin{table}
\label{tab4}
\centering
\caption{Dissociation of upper bound for N=5, temperatures are in the unit of $T_c$ by using thermal energy effect criteria.}
{\begin{tabular}{@{}cccc@{}} \toprule
$State$$\Downarrow$ & eB=0.3 $GeV^{2}$ & eB=0.5 $GeV^{2}$ & eB=0.7 $GeV^{2}$\\ \colrule
$J/\psi$ & 2.4310 & 2.4209 & 2.4107\\ \colrule
$\Upsilon$ & 2.7010 & 2.6909 & 2.6799\\ \botrule
\end{tabular}}
\end{table}
\begin{equation}
\label{eq36} 
d_{1}-2\beta (\delta +2)=2\alpha (\alpha_{1}^{(2)}+\alpha_{2}^{(2)}  )
\end{equation}
\begin{equation}
\label{eq37}
(d_{1}-2\beta \delta)\alpha_{1}^{(2)} \alpha_{2}^{(2)}=2\delta (\alpha_{1}^{(2)}+\alpha_{2}^{(2)})
\end{equation}
\begin{equation}
\label{eq38}
{d_{1}-2\beta (\delta +1)}(\alpha_{1}^{(2)}+\alpha_{2}^{(2)}  )=4\alpha(\alpha_{1}^{(2)} \alpha_{2}^{(2)})+2(2\delta +1)
\end{equation}
The energy eigen value for $E_{2l}$ is:
\begin{equation}
\label{eq39}
E_{2l}=\sqrt{\frac{a}{2\mu_{Q\bar{Q}}}}(N+2l+4)+C-\frac{b^{2}}{4(\sqrt{a})^{2}}
\end{equation}
Hence, with the repetition of the Iteration method, the exact energy eigen value for the quarkoniun states depending upon the temperature and magnetic field in the N-dimensional space becomes:
\begin{equation}
\label{eq40}
E_{nl}^{n}=\sqrt{\frac{a}{2\mu_{Q\bar{Q}}}}(N+2l+2n)+C-\frac{b^{2}}{4(\sqrt{a})^{2}}
\end{equation} $n=0, 1, 2, 3........$.
\begin{figure*}
\centering
    \vspace{3mm}   
    \includegraphics[height=8cm,width=14cm]{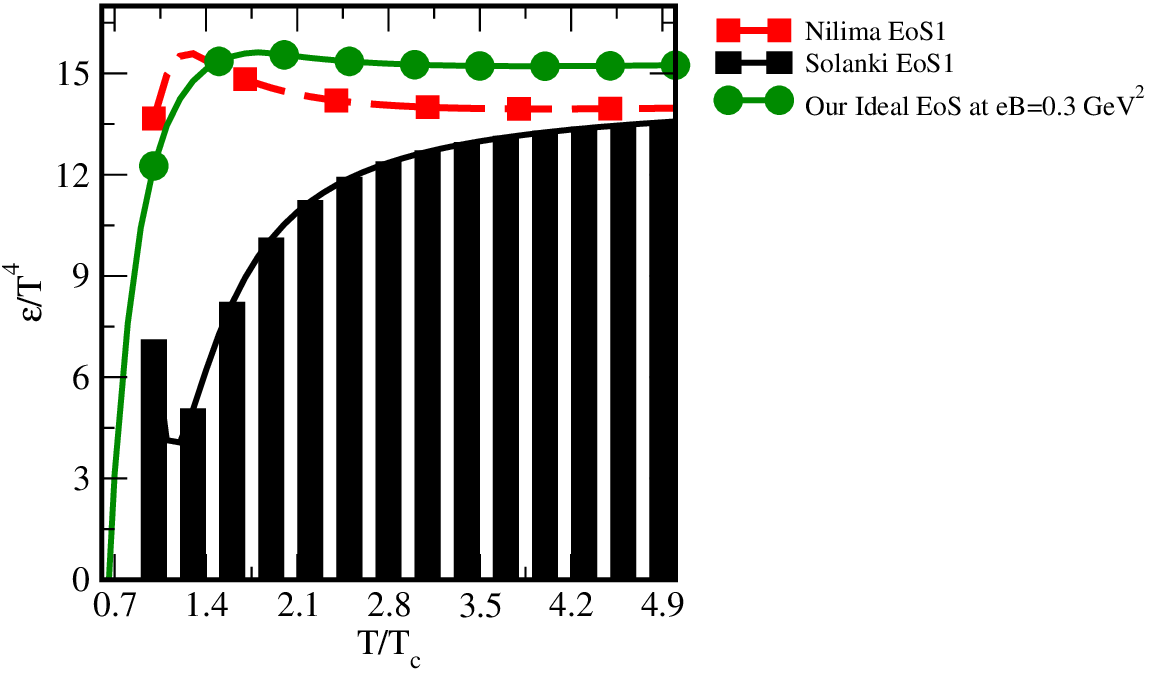}
\vspace{2cm}
\caption{Variation of $\frac{\epsilon}{T^{4}}$ as a function of $\frac{T}{T_{c}}$ for our ideal EoS at eB=0.3 Ge$V^{2}$ and also compared with the published results of Nilima EoS1 \cite{56} and Solanki EoS1 \cite{57}.}
\label{fig.7}
 \vspace{3cm} 
\end{figure*}
\begin{table}
\label{tab5}
\centering
\caption{Dissociation temperature for eB=0.3 Ge$V^{2}$, temperatures are in the unit of $T_c$ by using dissociation energy criteria.}
{\begin{tabular}{@{}cccc@{}} \toprule
$State$$\Downarrow$ & N=3 & N=4 & N=5\\ \colrule
$J/\psi$ & 2.1959 & 2.3090 & 2.4055\\ \colrule
$\Upsilon$ & 2.2965 & 2.4711 & 2.6079\\ \botrule
\end{tabular}}
\end{table}
\begin{table}
\label{tab6}
\centering
\caption{Dissociation temperature for N=5, temperatures are in the unit of $T_c$ by using dissociation energy effect criteria.}
{\begin{tabular}{@{}cccc@{}} \toprule
$State$$\Downarrow$ & eB=0.3 $GeV^{2}$ & eB=0.5 $GeV^{2}$ & eB=0.7 $GeV^{2}$\\ \colrule
$J/\psi$ & 2.4055 & 2.3949 & 2.3874\\ \colrule
$\Upsilon$ & 2.6079 & 2.5984 & 2.5913\\ \botrule
\end{tabular}}
\end{table}

\section{Quasi-particle model and Debye mass}
In the quasi-particle description, the system of the interacting particles supposed to be non-interacting or in other words weakly interacting by means of the effective fugacity\cite{45} or with the effective mass\cite{46,47}.
Nambu-Jona-Laisino (NJL) and Ploylov Nambu-Jona-Laisino (PNJL) quasi-particles models\cite{48} self-consistent quasi-particles model\cite{49} etc, include the effective masses. Here we considered the effective fugacity quasi-particle model (EQPM), in the presence of eB, which interprets the QCD EoS as non interacting quasi-partons with effective fugacity parameter $z_{g}$ for gluons and $z_{q}$ for quarks encoding all the interacting effects taking place in the medium. The distribution function for quasi-gluons and the quasi-quarks/quasi-anti quarks\cite{50} are given in the presence of magnetic field as below:
\ba
f_{g/q}=\frac{f_{g/q}e^{-\beta E_p}}{1\mp f_{g/q}e^{-\beta E_p}}
\label{41}
\ea
To measure the effect of electric potential applied on the QGP, Debye mass played a major role and is gauge invariant and non-perturbative in nature. There are several model developed to explain the system of weakly interacting particles such as, effective fugacity model\cite{50}, NJL and PNJL model\cite{53}, self-consistent quasi-particles\cite{54} etc. In this work we have considered the EQPM which is extended in the presence of magnetic field and interprets the QCD EoS as non-interacting quasi-partons with $z_{q}$ and $z_{g}$ as fugacity parameter. So, the distribution function for quark/anti-quark is given below: 
\ba
f^{o}_{q}=\frac{z_{q} e^{-\beta \sqrt{p^{2}_{z}+m^{2}+2l\vert q_{f}eB\vert}}}{1+{z_{g} e^{-\beta \sqrt{p^{2}_{z}+m^{2}+2l\vert q_{f}eB\vert}}}}
\label{42}
\ea
Where, l is the landau quantum number, eB is the magnetic field. The effect of the magnetic field $B=B\hat{z}$ is taken along the $z$-axis. Since, the plasma contains both the charged and the quasi-neutral particles, hence it shows collective behavior. Debye mass is an important quantity to describe the screening of the color forces in the Hot QCD medium. Debye screening mass can be defined as the ability of the plasma to shield out the electric potential applied to it. In the studies\cite{52,53,54,55}, detailed definition of the Debye mass can be seen. To determine the Debye mass in terms of the eB, we start from the gluon-self energy as below:
\ba 
m^{2}_{D}=\Pi_{00}(\omega =p,\vert \overrightarrow{p} \vert\longrightarrow 0)
\label{43}
\ea
According to the\cite{56}, gluon self energy was modified as:
\begin{equation}
\Pi_{00}(\omega =p,\vert \overrightarrow{p} \vert\longrightarrow 0)=\frac{g^{2}\vert eB \vert}{2 \pi^{2}T} \int^{\infty}_{0} dp_{z} f^{0}_{q}(1-f^{0}_{q})
\label{44}
\end{equation}
\begin{figure*}
\centering
    \vspace{3mm}   
    \includegraphics[height=8cm,width=14cm]{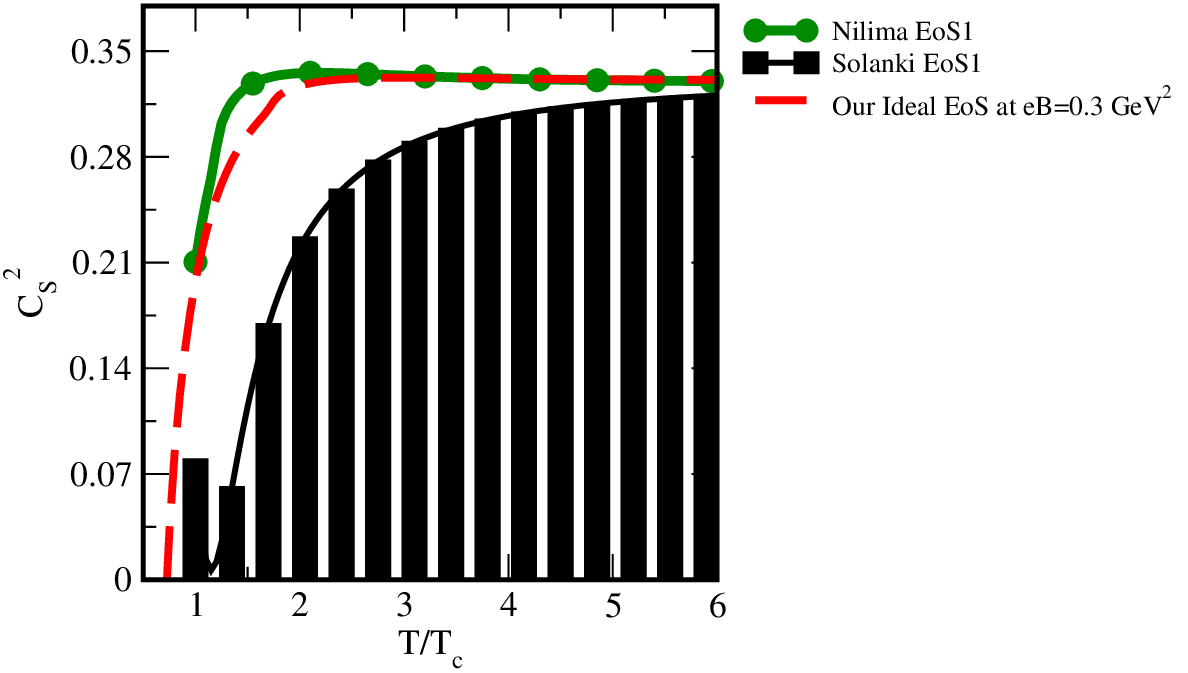}
\vspace{2cm}
\caption{Variation of $c_{s}^{2}$ as a function of $\frac{T}{T_{c}}$ for our ideal EoS at eB=0.3 Ge$V^{2}$ and also compared with the published results of Nilima EoS1 \cite{56} and Solanki EoS1 \cite{57}.}
\label{fig.8}
 \vspace{3cm} 
\end{figure*}
Thus, Debye mass for quarks using the distribution function defined by Eq.(\ref{42}) as given below:
\ba
m^{2}_{D}=\frac{4\alpha}{\pi T}|eB|\int^{\infty}_{0} dp_{z} f^{0}_{q}(1-f^{0}_{q})
\label{45}
\ea
Since, the magnetic field has no effect on the gluon, gluonic contribution to the Debye mass, it will remains intact/unchanged. The other approach to obtain the Debye mass is the kinetic theory approach. Both these approaches provide similar results for the Debye mass in the presence of eB. So, the Debye mass for $N_{f}$=3 and $N_{c}$=3 will be:
\ba
m^{2}_{D}= 4 \alpha \left(\frac{6T^{2}}{\pi} PolyLog[2,z_{g}]+\frac{3eB}{\pi} \frac{z_{q}}{1+z_{q}}\right) 
\label{46}
\ea
The Debye mass for the ideal EoS [$z_{q,g}$=1] representing non-interacting quarks and gluons becomes:
\ba
m^{2}_{D}= 4 \pi \alpha \left( T^{2}+\frac{3eB}{2\pi^{2}}\right)
\label{47}
\ea 

\section{Binding energy (B.E.) of Quarkonium state in N-dimensional space}
The binding energy of quarkonium state such as Bottomonium and Charmonium have been studied in this section. By using AEIM method, the exact energy eigen values in N-dimensional space becomes:
\begin{equation}
\label{eq48}
B.E.=E_{nl}^{n}=\sqrt{\frac{a}{2\mu_{Q\bar{Q}}}}(N+2l+2n)+C-\frac{b^{2}}{4(\sqrt{a})^{2}}
\end{equation} 
In this equation $(n= 0, 1, 2, 3....)$ corresponding to the state of quarkonia.

\section{Mass Spectra of Quarkonium state in N-dimensional space}
The mass spectra of heavy quarkonia can be calculated by using the relation given below:
\begin{equation}
\label{eq49}
M=2m_{Q}+B.E.
\end{equation}
Here, mass spectra is equal to the sum of the energy eigen values and twice  of the quark-antiquark mass. Substituting the values of $E_{nl}^{n}$ in the Eq.(\ref{eq49}) we get,
\begin{equation}
\label{eq50}
M=2m_{Q}+\sqrt{\frac{a}{2\mu_{Q\bar{Q}}}}(N+2l+2n)+C-\frac{b^{2}}{4(\sqrt{a})^{2}}
\end{equation}
\begin{figure*}
\centering
    \vspace{3mm}   
    \includegraphics[height=7cm,width=11cm]{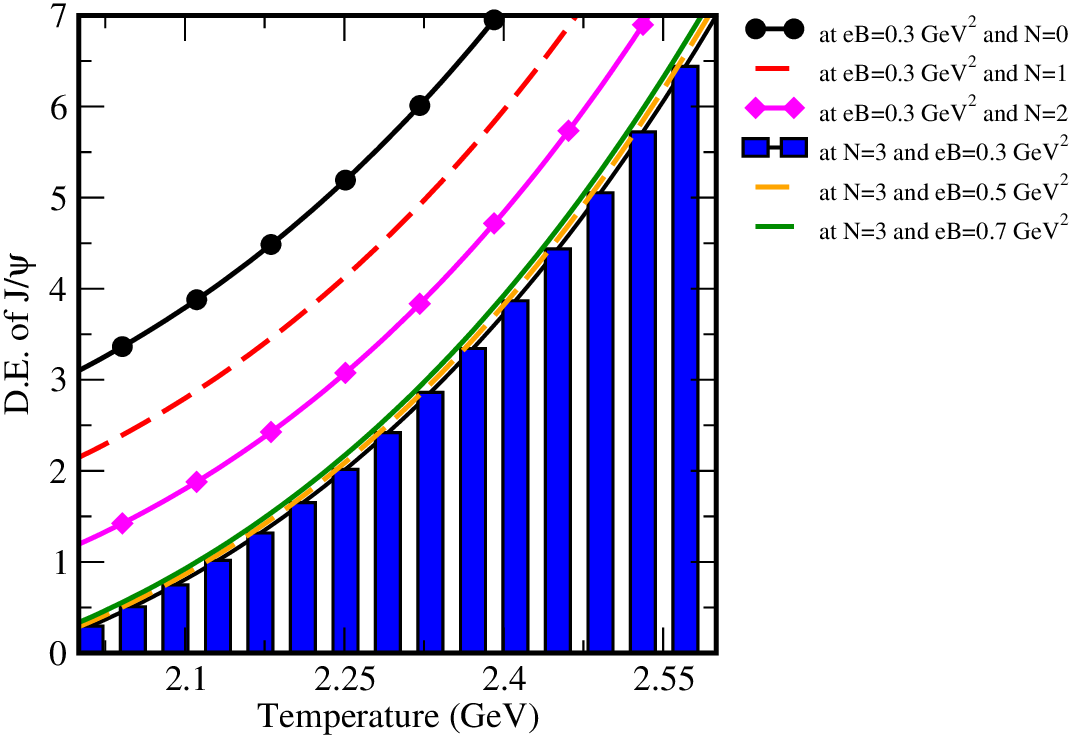}
     \vspace{2cm}  
    \caption{The variation of dissociation energy (D.E.) of J/$\psi$ with temperature at different values of magnetic field (eB) and dimensionality number (N).}
\label{fig.9}
 \vspace{3cm}  
\end{figure*}
Where, $m_{Q}$ is the mass of quarkonium state, l is the angular momentum quantum number and $\mu_{Q\bar{Q}}$ is the reduced mass.\\

\section{QCD EoS in the presence of magnetic field}
The EoS for the quark-matter is an important findings in relativistic nucleus-nucleus collisions, and the thermodynamical properties of matter are sensitive to EoS. The EoS which is defined as a function of plasma parameter $(\Gamma )$\cite{39} is,
\begin{eqnarray}
\label{eq51}
\epsilon_{QED}=\left [ \frac{3}{2}+\mu_{ex}(\Gamma )  \right ]nT
\end{eqnarray}
The ratio of average potential energy to average kinetic energy is known as plasma parameter. Now, let as assumed that $\Gamma \ll 1$ and is given by:
\begin{eqnarray}
\label{eq52}
\Gamma\equiv \frac{<PE>}{<KE>}= \frac{Re[V(r,T)]}{T}
\end{eqnarray}
But after inclusion of relativistic and quantum effects, the EoS which has been already obtained in the $\Gamma$ can be written as:
\begin{eqnarray}
\label{eq53}
\epsilon =\left ( 3+\mu_{ex}(\Gamma) \right )nT
\end{eqnarray}
The scaled-energy density is written as in terms of ideal contribution\cite{56,57} given below as:
\begin{eqnarray}
\label{eq54}
e(\Gamma)\equiv \frac{\epsilon }{\epsilon_{SB } }=1+\frac{1}{3}\mu_{ex}(\Gamma)
\end{eqnarray}

\begin{table*}
\label{tab7}
\centering
\caption{Comparison of the mass spectra for $J/\psi$ and $\Upsilon$ obtained in the present work at N=$1$ with the theoretical and experimental data.}
{\begin{tabular}{@{}cccccc@{}} \toprule
$State$ $\Downarrow$ & eB=0.3 $GeV^{2}$ & eB=0.5 $GeV^{2}$ & eB=0.7 $GeV^{2}$ & Solanki\cite{57} & Exp.mass\cite{Tanabashi}\\ \colrule
J/$\psi$ & 3.1191 & 3.0590 & 3.0161 & 3.060 & 3.096\\ \colrule
$\Upsilon$ & 10.4314 & 10.3824 & 10.3211 & 9.200 & 9.460\\ \botrule
\end{tabular}}
\end{table*}
Where, $\epsilon_{SB}$ is:
\begin{eqnarray}
\label{eq55}
\epsilon_{SB}\equiv (16+21N_{f}/2)\pi^{2}T^{4}/30
\end{eqnarray}\\
Here, $N_{f}$ is the number of flavor of quarks and gluons, and we also consider two-loop level QCD running coupling constant ($\alpha$) in $\overline{MS}$ scheme\cite{56,57},
\begin{eqnarray}
\label{eq56}
g^{2}(T)\approx 2b_{0}ln\frac{\bar{\mu}}{\Lambda_{\overline{MS}} }\left ( 1+\frac{b_{1}}{2b_{0}^{2}}\frac{ln\left ( 2ln\frac{\bar{\mu}}{\Lambda_{\overline{MS}} } \right )}{ln\frac{\bar{\mu}}{\Lambda_{\overline{MS}} }} \right )^{-1}
\end{eqnarray}
Here, $b_{0}=\frac{33-2N_{f}}{48\pi^{2}}$ and $b_{1}=\frac{153-19N_{f}}{384\pi^{4}}$. In $\overline{MS}$ scheme, $\Lambda_{\overline{MS}}$ and $\bar{\mu}$ are the renormalization scale and the scale parameter respectively. The dependency of $\Lambda_{\overline{MS}}$ is:
\begin{eqnarray}
\label{eq57}
\bar{\mu}exp(\gamma_{E}+c)=\Lambda _{\overline{MS}}(T)\nonumber\\
\Lambda _{\overline{MS}}(T)exp(\gamma_{E}+c)=4\pi\Lambda_{T}.
\end{eqnarray}
Here, $\gamma_{E}=0.5772156$ and $c=\frac{N_{c}-4N_{f}ln4}{22N_{c}-N_{f}}$\cite{56,57}. After using the above relation, first we calculated the energy density $\epsilon_{T}$ from Eq.\ref{eq54} and using the thermodynamical relation:
\begin{eqnarray}
\label{eq58}
\epsilon =T\frac{dp}{dT}-P
\end{eqnarray}
We calculated the pressure as:
\begin{eqnarray}
\label{eq59}
\frac{P}{T^{4}}=\left ( \frac{P_{0}}{T_{0}}+3a_{f}\int_{T_{0}}^{T}d\tau \tau^{2}\epsilon (\Gamma (\tau ))  \right )/T^{3}
\end{eqnarray}
Here, $P_{0}$ is the pressure at some reference temperature $T_{0}$. Now, the speed of sound is calculated by using the realtion as:
\begin{eqnarray}
\label{eq60}
c_{s}^{2}=\left ( \frac{dP}{d\epsilon } \right )
\end{eqnarray}
All above thermodynamical properties are potential dependent, the potential is Debye mass dependent. Hence in that case, we invaded the problem by trading off the dependence on magnetic field, to a dependence on these thermodyanmic properties of matter.
\begin{figure*}
    \vspace{5mm}   
    \includegraphics[height=6cm,width=8cm]{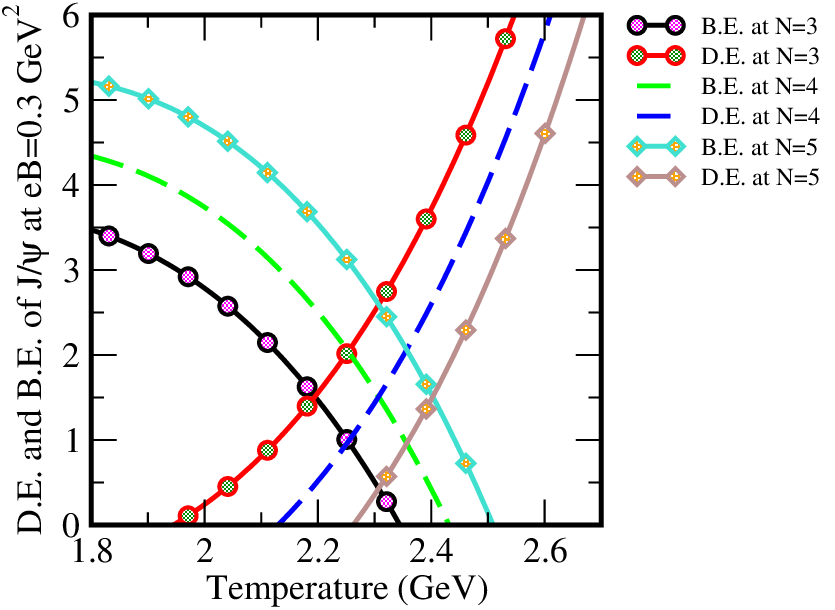}
    \hspace{8mm}
    \includegraphics[height=6cm,width=8cm]{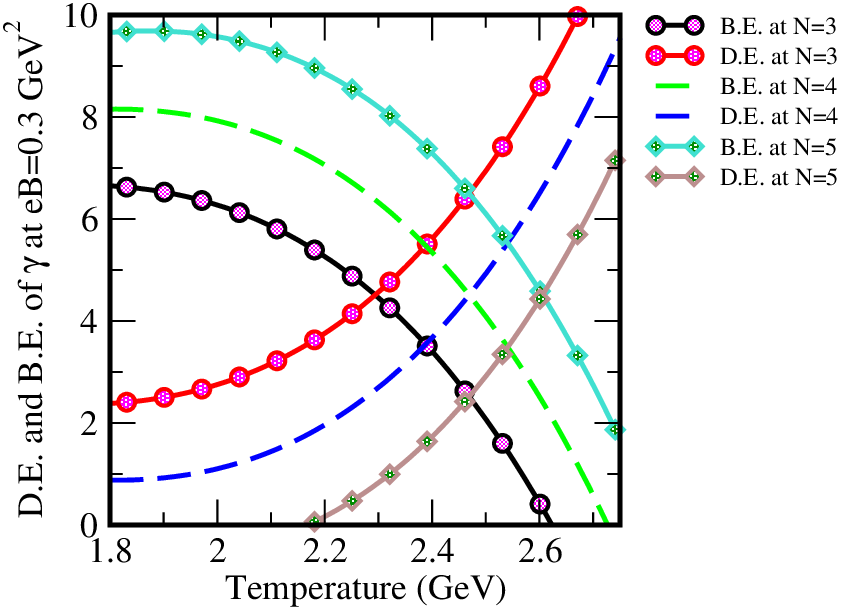}
\vspace{2cm}
\caption{Shows the variation of dissociation energy (D.E.) and binding energy (B.E.) of $J/\psi$ (left panel) and $\Upsilon$ (right panel) with temperature at different values of dimensionaity number (N) and fixed value of magnetic field (eB).}
\label{fig.10}
 \vspace{3cm} 
\end{figure*}
\section{Dissociation energy (D.E.) of Quarkonium state in N-dimensional space}
The most precious quantity to notice the vanishing of bound state is the Dissociation energy. The D.E. of heavy quarkonia can be calculated by using the relation given below:
\begin{equation}
\label{eq61}
D.E.=E_{dissociation}^{N,l}\equiv 2m_{Q}+\frac{\sigma}{m_{D}}-B.E. 
\end{equation}
After introducing the value of binding energy in the above expression, we get the final expression of D.E. in N-dimensional space as:
\begin{multline}
\label{eq62}
D.E.=E_{dissociation}^{N,l}\equiv2m_{Q}+\frac{\sigma}{m_{D}}\\-\left[\sqrt{\frac{a}{2\mu_{Q\bar{Q}}}}(N+2l+2n)+C-\frac{b^{2}}{4(\sqrt{a})^{2}} \right]
\end{multline}

\section{Results and Discussion}
\label{RD}
In this analysis, we have taken fixed value of critical temperature ($T_{c}$=197 MeV) throughout the manuscript and various quantities such as binding energy (B.E.), dissociation temperature and the mass spectra of the quarkonia has been studied with finite values of magnetic field. The variation of the potential with distance (in fm) at fixed value of temperature (T=250 MeV) for different values of magnetic field (eB=0.3, 0.5 and 0.7 Ge$V^{2}$) is shown in figure \ref{fig.1}. We clearly noticed that in figure \ref{fig.1}, if we increases the values of magnetic field then the variation of potential also increases.\\
Figure \ref{fig.2} and \ref{fig.3} shows the variation of binding energy of J/$\psi$ and $\Upsilon$ with temperature. From figure \ref{fig.2}, we can deduce that the values of the binding energy of J/$\psi$ (in left panel) and $\Upsilon$ (in right panel) decrease with temperature for different values of the magnetic field (eB=0.3, 0.5 and 0.7 Ge$V^{2}$) at fixed value of dimensionality number (N=3). The effect of dimensionality number on the binding energy for the quarkonium state J/$\psi$ (in left panel) and $\Upsilon$ (in right panel) with temperature has been shown in figure \ref{fig.3}. If the value of dimensionality number increases, the binding energy of quarkonium states becomes higher at the fixed value of magnetic field (eB=0.3 Ge$V^{2}$).\\
Figure \ref{fig.4}, shows the variation of mass spectra of heavy quarkonia with temperature for 1S state of charmonium J/$\psi$ (left panel) and 1S state of bottomonium $\Upsilon$ (right panel) for different values of magnetic field (eB=0.3, 0.5 and 0.7 Ge$V^{2}$) at fixed value of dimensionality number (N=1). We observed that, if we increases the value of magnetic field (at fixed value of mass of ground state of quarkonium i.e., $m_{J/\psi}$=1.5 GeV and $m_{\Upsilon}$=4.5 GeV) the variation of mass spectra decreases. Figure \ref{fig.5}, also shows the variation of mass spectra of heavy quarkonia with temperature for 1S state of charmonium J/$\psi$ (left panel) and 1S state of bottomonium $\Upsilon$ (right panel) for different values of dimensionality number (N=0, 1 and 2) at fixed value of magnetic field (eB=0.7 Ge$V^{2}$). We observed that, if we increase the values of dimensionality number, the variation of mass spectra also increases. We also compared the value of mass spectra for J/$\psi$ and $\Upsilon$ obtained in the present work at different values of magnetic field with the other calculated theoretical and experimental values of mass spectra as shown in table \ref{tab7}.\\
The dissociation temperature for real binding energies can be obtained by using thermal energy. According to the references~\cite{26,38} it is not necessary to have zero binding energy for dissolution of the quarkonium states. When binding energy $(E_{bin}\le T$) of quarkonium state is weakly bonded, it dissociates by means of thermal fluctuations. The quarkonium state is also said to be dissociated when 
$2B.E\leq\mathrm{\Gamma }(T)$, $\mathrm{\Gamma}(T)$ is thermal width of respective quarkonium states. Hence, there are two ways to calculate dissociation point of quarkonia. The lower bound of dissociation using mean thermal energy can be written as:
\begin{equation}
\label{eq63}
\sqrt{\frac{a}{2\mu_{Q\bar{Q}}}}(N+2l+2n)+C-\frac{b^{2}}{4(\sqrt{a})^{2}}=3(T_D) 
\end{equation}
and for upper bound the expression is, as follow:
\begin{equation}
\label{eq64}
\sqrt{\frac{a}{2\mu_{Q\bar{Q}}}}(N+2l+2n)+C-\frac{b^{2}}{4(\sqrt{a})^{2}}=(T_D) 
\end{equation}
Where, $\mu_{Q\bar{Q}}$=$\frac{m_{c}}{2}$ and $\frac{m_{b}}{2}$.
The dissociation temperatures for the J/$\psi$ and $\Upsilon$ has been given in the tables \ref{tab1}, \ref{tab2}, \ref{tab3} and \ref{tab4}. Lower bound of dissociation temperature has been shown in table \ref{tab1} (at different values of N) and \ref{tab2} (at different values of magnetic field). Whereas the tables \ref{tab3} (at different values of N) and \ref{tab4} (at different values of magnetic field) shows the different values of dissociation temperatures for upper bound. In general, the dissociation temperature decreases with the magnetic field and increases with dimensionality number.\\
\begin{figure*}
    \vspace{5mm}   
    \includegraphics[height=6cm,width=8cm]{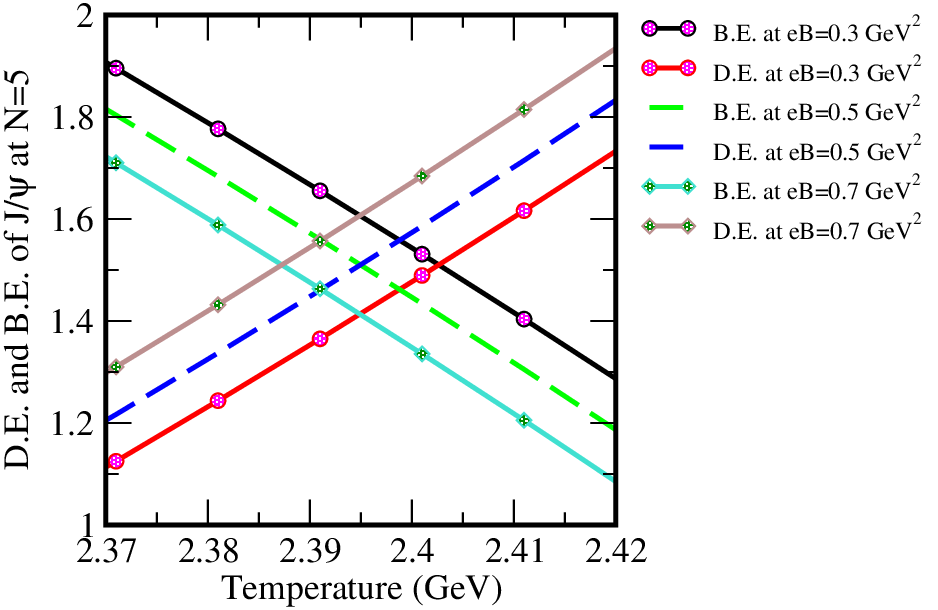}
    \hspace{8mm}
    \includegraphics[height=6cm,width=8cm]{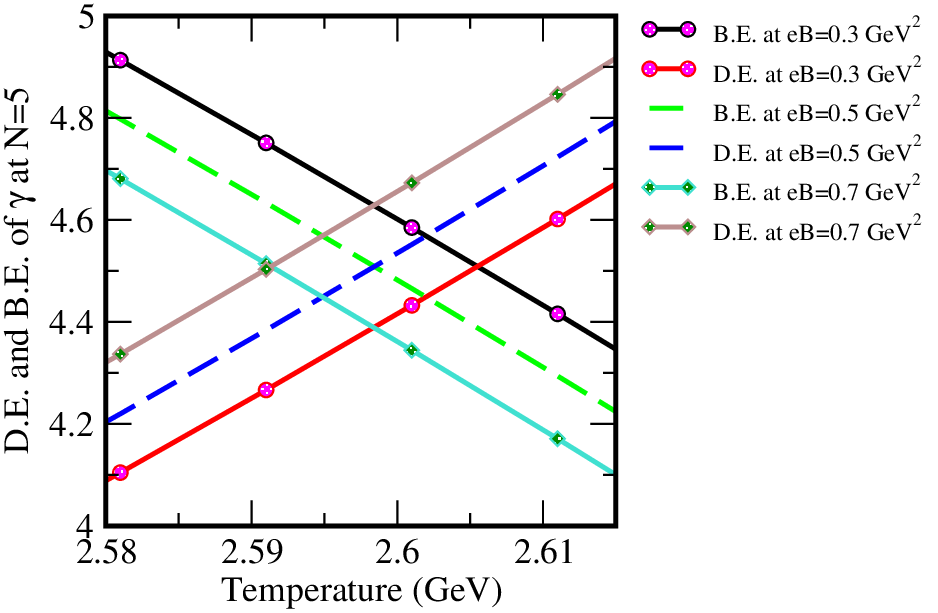}
\vspace{1cm}
\caption{Shows the variation of dissociation energy (D.E.) and binding energy (B.E.) of $J/\psi$ (left panel) and $\Upsilon$ (right panel) with temperature at different values of magnetic field (eB) and fixed value of dimensionality number (N).}
\label{fig.11}
 \vspace{1cm} 
\end{figure*}
The variation of dissociation energy (D.E.) of J/$\psi$ has been shown in the figure \ref{fig.9} with the temperature for different values of magnetic field and dimensionality number. It has been deduced that the D.E. of the quarkonium states with the magnetic field increases, but with the dimensionality number it decreases. In this manuscript, we have applied one more way to calculate the dissociation temperature with the help of dissociation energy (D.E.) and binding energy. The variation of B.E. and D.E. of the J/$\psi$ in the figure \ref{fig.10} and \ref{fig.11} (left panel) and $\Upsilon$ in the figure \ref{fig.10} and \ref{fig.11} (right panel) with temperature for different values of dimensionality number figure \ref{fig.10} and for different values of magnetic field figures \ref{fig.11} has been shown respectively. From the figure \ref{fig.10} and \ref{fig.11} we have examined the dissociation temperature (by intersection point of the D.E. and B.E. of the quarkonia) of the J/$\psi$ and $\Upsilon$ for different values of eB and N, and the values of dissociation temperature is given in the table \ref{tab5} and \ref{tab6} respectively.\\
The thermodynamical properties of quark matter plays a significant role in the study of QGP and also provide useful information about the strange quark-matter. In figure \ref{fig.6}, \ref{fig.7} and \ref{fig.8} we have plotted the variation of pressure $(\frac{P}{T^{4}})$, energy density $\epsilon$ and speed of sound $(C_{s}^{2})$ with temperature (T/$T_{c}$) at eB=0.3 Ge$V^{2}$ for ideal EoS with 3 flavor QGP. The obtained results also compared with the results of Nilima EoS1~\cite{56} and Solanki EoS1 \cite{57} and they were found to be consistent with \cite{56,57}.\\

\section{Conclusion}
\label{CO}
We have considered the medium modified form of Cornell potential at finite values of magnetic field and dimensionality number. To reach end, we considered magnetic field dependent quasi-particle Debye mass for the study of dissociation pattern of quarkonia. Real part of medium modified form of Cornell potential has been used for solving schrodinger equation to obtain binding energy of quarkonia in N-dimensional space. We observed that binding energy and mass spectra decreases with increasing the values of eB. However, binding energy and mass spectra tends to get higher with increasing value of N. The variation of dissociation energy with the temperature for different values of eB and N has been shown in the figure \ref{fig.9}. It has been seen that the dissociation energy of the quarkonium states with the magnetic field increases but with the dimensionality number decreases.\\
We applied an another way to calculate the dissociation temperature with the help of dissociation energy and binding energy. The variation of B.E. and D.E. of the states of quarkonia with temperature for different values of dimensionality number and for different values of magnetic field has been shown respectively. From these, we have examined the dissociation temperature at the intersection point of the D.E. and B.E. of the quarkonia for different values of eB and N.\\
In conclusion, the dissociation temperature of heavy quarkonia decreases with magnetic field and increases with dimensionality number. We have also extended this work, after calculating the thermodynamical properties of QGP (i.e., pressure, energy density and speed of sound) using the parameters eB and N. In future, we may extend this work to calculate the nucleus-nucleus suppression with the latest determined value of $\sqrt{s_{NN}}$. Also we can study about the survival probability or nuclear modification factor of different quarkonium states w.r.t eB, dimensionality number, transverse momentum, centrality, and rapidity which is the key point to quantify various properties of the medium produced during Heavy Ion Collisions (HICs) at LHC and RHIC.\\

\section{Acknowledgments}
One of the authors, VKA acknowledges the Science and Engineering Research Board (SERB) Project No. EEQ/2018/000181 New Delhi for the research support in basic sciences.

\end{document}